\documentclass[preprint,12pt,showpacs]{revtex4}

\usepackage[brazil, english]{babel}
\usepackage[utf8]{inputenc}
\usepackage{graphicx}
\usepackage{epsfig}
\usepackage{amssymb}
\usepackage{amsmath}

\graphicspath{%
    {converted_graphics/}% inserted by PCTeX
    {/}% inserted by PCTeX
}
\begin{document}

%\date{}
\title{Odd spin glueball masses and the odderon Regge trajectories from the holographic hardwall model}
\author{Eduardo Folco Capossoli$^{1,2,}$}
\email[Eletronic address: ]{educapossoli@if.ufrj.br}
\author{Henrique Boschi-Filho$^{1,}$}
\email[Eletronic address: ]{boschi@if.ufrj.br}  
\affiliation{ $^1$Instituto de F\'{\i}sica, Universidade Federal do Rio de Janeiro, 21.941-972 - Rio de Janeiro-RJ - Brazil \\
 $^2$Departamento de F\'{\i}sica, Col\'egio Pedro II, 20.921-903 - Rio de Janeiro-RJ - Brazil}

\begin{abstract}
We use the holographic hardwall model to calculate the masses of light glueball states with odd spin and $P= C= -1$ associated with odderons. Considering Dirichlet and Neumann boundary conditions we obtain expressions for the odderon Regge trajectories consistent with those calculated using other approaches.
\end{abstract}


\maketitle

%\begin{multicols}{2}

\section{Introduction}

For mesons and baryons, there is an approximate relation between the total angular momenta $(J)$ and the square of the masses $(m)$. This relation, known as Regge trajectories has the form 
\begin{equation}\label{regge}
J(m^2) \simeq \alpha_0 + \alpha 'm^2 ,
\end{equation}

\noindent where $\alpha_0$ and $\alpha '$ are constants. Analogously one can find Regge trajectories for odd spin glueballs with $P= C= -1$, which are related to the odderon.

The Regge trajectories for the odderon were obtained by Llanes-Estrada {\sl et. al.} \cite{LlanesEstrada:2005jf} using two different methods. The first one is based on a relativistic many-body (RMB) formulation that gives 
\begin{equation}\label{l1}
J_{RMB}(m^2) = -0.88 + 0.23 m^2 ,
\end{equation}

\noindent where the masses are expressed in GeV throughout this article. The second method is based on a nonrelativistic constituent model (NRCM) resulting in 
\begin{equation}\label{l2}
J_{NRCM}(m^2) = 0.25 + 0.18 m^2 .
\end{equation}
Interesting studies of the odderon in gauge/string dualities were presented in Refs. 
\cite{Brower:2008cy,Avsar:2009hc}. 

In this work we obtain the masses of odd spin glueballs from the holographic hardwall model and derive the corresponding Regge trajectories for the odderon. We find results compatible with those above, given by Eqs. (\ref{l1}) and (\ref{l2}).

Since its conception, quantum chromodynamics (QCD) has been used as the standard theory to explain the phenomenology of strong interactions. As a consequence of asymptotic freedom, the coupling of strong interactions decreases when the energy of the process increases. This result is obtained using perturbation theory and is valid only for small couplings $(g < 1)$. Extrapolating this result to low energies one obtains strong coupling $(g > 1)$ outside the perturbative regime. Regge trajectories are an example of nonpertubative behavior of strong interactions difficult to model  using QCD. 

The anti-de Sitter/conformal field theory (AdS/CFT) correspondence \cite{Maldacena:1997re, Gubser:1998bc, Witten:1998qj, Witten:1998zw, Aharony:1999ti, Petersen:1999zh} brought new perspectives for string and quantum field theories since it relates $SU(N)$ supersymmetric and conformal Yang-Mills field theory for $N \rightarrow \infty$, in flat Minkowski spacetime with $3 + 1$ dimensions, with a string theory in a curved 10-dimensional spacetime, the AdS$_{5} \times S^5$ space. In the supergravity approximation of string theory in this space one can relate both theories through \cite{Witten:1998qj, Gubser:1998bc} 
\begin{equation}\label{presc}
Z_{\rm CFT}[\varphi_o]= \Biggl <  \exp \Biggl ( \int_{\partial \Omega} d^4 x \; {\cal O} \varphi_o \Biggr ) \Biggr > = \int_{\varphi_o} D \varphi \exp (-I_s(\varphi)),
\end{equation}

\noindent where $\varphi$ is a non-normalizable supergravity field, $ I_s(\varphi)$ is the corresponding on shell supergravity action, $\varphi_o$ is the value of $\varphi$ at the boundary $\partial \Omega$, and $\cal O$ is the associated operator of the conformal field theory (CFT). From this equation, one can obtain four-dimensional correlation functions, for instance  
\begin{equation}\label{corr}
\bigl <  {\cal O}(x) {\cal O}(y) \bigr > = \frac {\delta^2 Z_{\rm CFT}[\varphi_o]}{\delta \varphi_o(x) \delta \varphi_o(y)}\Bigg|_{\varphi_o = 0 }.
\end{equation}

In particular, the scalar glueball $0^{++}$ is represented by the operator ${\cal O}_{4} = F^2$ associated with a dilaton in the AdS$_{5} \times S^5$ space.

\section{Odd spin glueball masses in the Hardwall model} 

Glueballs are characterized by $J^{PC}$, where $J$ represents the total angular momentum, $P$ defines how a state behaves under spatial inversion ($P$-parity), and $C$ shows the behavior of a state under charge conjugation ($C$-parity). 

In this paper we are interested in glueballs in the $P=-1$ and $C= -1$ sector with odd spins $ J \geq 1$, which are associated with a particle called the \emph {odderon}. The concept of the odderon emerged in the 1970s \cite{Lukaszuk:1973nt} within the context of asymptotic theorems, reappearing later in perturbative QCD \cite{Kwiecinski:1980wb, Bartels:1999yt}. The odderon has also been linked, for instance, to the color glass condensate \cite{Hatta:2005as}. Although the odderon has not been detected so far, it is regarded as a crucial test of QCD \cite{Avila:2006wy, Hu:2008zze}. The odderon is a bound state of three gluons, without color, which represents a singularity in the complex plane $J$, close to $1$, in the odd-under-crossing amplitude $F_-(s,t)$ \cite{Avila:2006wy}. 

%%%

The best experimental evidence for the odderon occurred in 1985 at ISR CERN. A difference between differential cross sections for ${pp}$ and ${p\bar{p}}$ in the dip-shoulder region $1.1 < |t| < 1.5 $ GeV$^{2}$ at $\sqrt{s} = 52.8$ GeV was measured, but these results were not confirmed \cite{Avila:2006wy}. 
There are two more evidences related to the nonperturbative odderon, that is, the change of shape in the polarization in $\,\,\pi^{-}p \rightarrow \pi^{0}n\,\,$ from $p_L = 5$ GeV/$c$ \cite{Hill:1973bq, Bonamy:1973dz} to $p_L = 40$ GeV/$c$ \cite{Apokin:1982kw} and a strange structure seen in the $UA4/2$ $dN/dt$ data for $pp$ scattering at $\sqrt{s} = 541$ GeV, namely a bump centered at $ |t| = 2 \times 10^{-3} $ GeV$^2$ \cite{Augier:1993sz}. Some other experiments to detect the odderon were proposed for the HERA \cite{Brodsky:1999mz} and recently for the LHC CERN  
through the study of coherent hadron-hadron interactions \cite {Goncalves:2012cy}. 
In this article, we are dealing with the nonperturbative odderon that is related with glueball states via its Regge trajectory, although the Regge trajectory of the odderon is not yet well understood. For reviews see, for instance, \cite{Mathieu:2008me, Ewerz:2003xi}.

%%%%

The AdS/CFT correspondence cannot be used directly as a tool for the study of hadrons because the dual theory is a supersymmetric conformal theory that is very different from QCD. However, it was noticed that the energy $E$ of a process in the $4d$ theory is related to the radial coordinate $z$ in AdS space as 
\begin{equation}\label{z_rad}
E \propto  \frac{1}{z} .
\end{equation}

This motivated the holographic hardwall model proposed by Polchinski and Strassler \cite{Polchinski:2001tt, Polchinski:2002jw} to calculate the scattering of glueballs in four dimensions using a dilaton field in AdS$_5 \times S^5$ space. The works \cite{BoschiFilho:2002ta, BoschiFilho:2002vd} introduced a cutoff at a certain value $z_{\rm max}$ of the $z$ coordinate and considered an AdS slice in the region  $0 \leq z \leq z_{\rm max}$. An immediate consequence of introducing a cutoff is the breaking of conformal invariance, so that particles on the four-dimensional boundary  acquire mass. Furthermore, one can associate the size of the AdS slice with the energy scale of QCD 
\begin{equation}\label{qcd}
z_{\rm max} = \frac{1}{\Lambda_{\rm QCD}} .
\end{equation}

%%%%%

Hadron masses can be determined using the hardwall model with a given mass scale (infrared cutoff). This can be used to build up Regge trajectories for the hadrons, as it was done in Refs. \cite{BoschiFilho:2005yh, deTeramond:2005su, Erlich:2005qh}.
One can note that the asymptotic behavior of these Regge trajectories is not linear. Despite this problem one can find approximate linear Regge trajectories for the first few light states of each hadronic branch.
It should be noticed that there is another holographic model that presents exact linear Regge trajectories: the softwall model \cite{Karch:2006pv}. The analysis presented in \cite{Karch:2006pv} for vector mesons can be extended to glueballs as was done in \cite{Colangelo:2007pt}.
However, despite the fact that the Regge trajectories are linear in this case, the glueball masses are too low compared with lattice data. 
So, in the following, we are going to use the hardwall model to study the odd spin glueball masses and to obtain the odderon Regge trajectories. 

%%%%

The hardwall model assumes an approximate duality between a string theory in an AdS$_5 \times S^5$ space with metric defined by 
\begin{equation}\label{met}
ds^2 = \frac{R^2}{z^2}(-dt^2 + d\vec x^2 + dz^2) + R^2 d\Omega^2_5 ,
\end{equation}

\noindent where $R$ is the AdS radius, and a pure Yang-Mills theory in four dimensions with symmetry group $SU (N)$ is in the large $N$ limit. In this model it is assumed that the AdS/CFT dictionary between supergravity fields in AdS$_5 \times S^5$ space and operators on the $4d$ boundary, as given by Eqs. (\ref{presc}) and (\ref{corr}), still holds after breaking the conformal invariance. This implies that the conformal dimension $\Delta$ of an operator $\cal O$ related to a $p$-form AdS$_5$ field with mass $m_5$ is given by \cite{Csaki:1998qr} (here and in the following we are disregarding excitations on the $S^5$ subspace) 
\begin{equation}\label{dim}
m^2_5R^2 = (\Delta - p)(\Delta + p - 4).
\end{equation}

In particular, the operator that describes the glueball $1^{--}$ is 
\begin{equation}\label{aco}
Sym Tr (\tilde F_{\mu \nu}F^2)
\end{equation}

\noindent with conformal dimension $\Delta = 6$. This operator is associated with the Ramond-Ramond tensor $C_{2,\sigma \lambda}$ described in a single $D3$-brane, by the action \cite{Brower:2000rp,Brower:2008cy} 
\begin{equation}\label{acao}
{\cal I} = \int d^4 x ~ \textrm {det} \left[G_{\sigma \lambda} + \exp^{- \frac{\phi}{2} } (B_{\sigma \lambda} + F_{\sigma \lambda}) \right] + \int d^4 x(C_0F \wedge F + C_2 \wedge F + C_4) .
\end{equation}

From this action one can obtain the equations of motion for the Ramond-Ramond field. With a suitable polarization choice $C_{2,\sigma \lambda}(x,z) = c_{\sigma \lambda} \phi(x,z)$ where $c_{\sigma \lambda}$ is a constant polarization tensor and $\phi(x,z)$ is a scalar field, it can be shown that these equations can be reduced to \cite{Constable:1999gb} 
\begin{equation}\label{esc}
\Biggl [ z^3 \partial_z \frac{1}{z^3} \partial_z + \eta^{\alpha \beta} \partial_{\alpha} \partial_{\beta} - \frac{m^2_5 R^2}{z^2} \Biggr ]\phi (x,z) = 0 ,
\end{equation}

\noindent where $\eta^{\alpha \beta}$ is the four-dimensional Minkowski metric.

 We use a plane wave ansatz in the four-dimensional space for the $0$-form field $\phi$ 
\begin{equation}\label{s_esc}
\phi (x,z) = A_{\nu , k} \exp^{-ip.x} z^2 J_{\nu}(u_{\nu ,\, k}\, z),
\end{equation}

\noindent where $A_{\nu , k}$ is a normalization constant, $J_{\nu}(y)$ is the Bessel function of order $\nu$ with $\nu = \sqrt{4 + m^2_5 R^2}$, and the discrete modes  $u_{\nu ,\, k}$ corresponding to the glueball masses will be calculated by imposing appropriate boundary conditions. Note that $ k = 1, 2 ,3, \dots,$ represent radial excitations of glueballs, but we will only consider in this paper the case $k=1$. 

It has been proposed in the literature \cite{deTeramond:2005su} that the glueball operator with spin $\ell$ could be obtained by the insertion of symmetrized covariant derivatives in the operator ${\cal O}_{4} = F^2$, such that ${\cal O}_{4+ \ell} = FD_{\{\mu_1 \cdots} {D_{\mu_\ell\}}}F$ with conformal dimension $\Delta = 4 + \ell$. This approach was used in Ref. \cite{BoschiFilho:2005yh} to calculate the masses of glueball states $0^{++}$, $2^{++}$, $4^{++}$, $6^{++}$, etc and to obtain the corresponding Pomeron Regge trajectory.

Here we are going to follow a similar approach for the glueball states $1^{--}$, $3^{--}$, $5^{--}$, $7^{--}$, $\dots$. The state $1^{--}$ is described by the operator $ {\cal O}_6 =  Sym Tr (\tilde F_{\mu \nu}F^2)$. Inserting covariant derivatives as described above, one obtains ${\cal O}_{6+ \ell} = Sym Tr (\tilde F_{\mu \nu} FD_{\{\mu_1 \cdots} {D_{\mu_\ell\}}}F)$ with $\Delta = 6 + \ell$ satisfying equations similar to (\ref{esc}) and (\ref{s_esc}) with a shift in the index of the Bessel function $\nu \rightarrow \nu = 4 + \ell$, where $\ell=J \geq 1$ is the spin of each state $1^{--}$, $3^{--}$, $5^{--}$, etc.

Following the approach of Ref. \cite{BoschiFilho:2005yh}, we impose Dirichlet and Neumann boundary conditions to calculate glueball masses within the hardwall model. For the Dirichlet boundary condition 
\begin{equation}\label{dir} 
\phi (z = z_{\rm max}) = 0
\end{equation}

\noindent one obtains from (\ref{s_esc}), the following masses: 
\begin{equation}\label{sdi}
u_{\ell,\, k}^{D} = \frac{\chi_{4 + \ell ,\, k}}{z_{\rm max}} = \chi_{4 + \ell ,\, k}\, \Lambda_{\rm QCD}; \; \; \; J_{4 + \ell}(\chi_{4 + \ell ,\, k})=0 .
\end{equation}

On the other hand, for the Neumann boundary condition 
\begin{equation}\label{neu}
\partial_z\phi |_{(z = z_{\rm max})} = 0 
\end{equation}

\noindent one gets 
\begin{equation}\label{sneu}
(\ell - 2) J_{4 + \ell} (\xi_{4 + \ell, \, k}) 
+ \xi_{4 + \ell,\, k} \, J_{3 + \ell} ( \xi_{4 + \ell, \, k})= 0 \,,
\end{equation}

\noindent and the masses are now
\begin{equation}\label{neu2}
u_{\ell,\, k}^{N} = \frac{\xi_{4 + \ell,\, k}}{z_{\rm max}} = \xi_{4 + \ell,\, k} \, \Lambda_{\rm QCD} .
\end{equation}

%%%

Using these boundary conditions we obtain glueball masses in the sector $P=C=-1$. 
We take the mass $u_{1,\, 1}$ of the state $1^{--}$ from the isotropic lattice ($3.24$ GeV) found in Refs. \cite{Meyer:2004jc, Meyer:2004gx} to fix $z_{\rm max} $ (and $\Lambda_{\rm QCD})$, and then calculate the other odd spin glueball masses 
$u_{\ell,\, 1}$ for the states $3^{--}$, $5^{--}$, $\dots$, using Eqs. (\ref{sdi}) and (\ref{neu2}), respectively, for the Dirichlet and Neumann boundary conditions. For instance, for the state 
$\ell^{--}$ with the Dirichlet boundary condition we have 
\begin{equation}\label{sol_3}
u_{\ell,\, 1}^D = \frac{\chi_{4 + \ell,\, 1}}{\chi_{5,\, 1}} {u_{1,\, 1}^D}\,,
\end{equation} 

\noindent so we get 4.09 GeV for the mass $u_{3,\, 1}^D$ of the $3^{--}$ state, etc. 
A similar calculation is done for the Neumann boundary condition.
%%% 
Our results are shown in Table I. We also show for comparison the values for these masses found in the literature \cite{LlanesEstrada:2005jf, Kaidalov:1999yd, Kaidalov:2005kz, Mathieu:2008pb, Chen:2005mg, Meyer:2004jc, Meyer:2004gx} using other methods.
%%%
Then from our results  we obtain different Regge trajectories for the odderon as discussed in the next section.

\begin{table}[h]
\caption{Glueball masses for states $J^{PC}$ expressed in GeV, with odd $J$ estimated using the hardwall model with Dirichlet and Neumann boundary conditions. The mass of $1^{--}$ is used as an input from the isotropic lattice \cite{Meyer:2004jc, Meyer:2004gx}. We also show other results from the literature for comparison.}
%\begin{ruledtabular}
%\label{t1}
\vspace{0.5 cm}
\centering
\begin{tabular}{|c|c|c|c|c|c|c|}
\hline
Models used &  \multicolumn{6}{c|}{Glueball states $J^{PC}$}  \\  
\cline{2-7}
 & $1^{ - - }$ & $3^{- - }$ & $5^{- - }$ & $7^{- - }$ & $9^{- - }$ & $11^{- - }$ \\
\hline \hline
Hardwall with Dirichlet b.c.                                   
&\, 3.24\, &\, 4.09 \,&\, 4.93 \,& \, 5.75 \,&\, 6.57 \,& \, 7.38 \, \\ \hline
Hardwall  with Neumann b.c.                                    
& 3.24 & 4.21 & 5.17 & 6.13 & 7.09 & 8.04      \\ \hline
Relativistic many body \cite{LlanesEstrada:2005jf}             
& 3.95 & 4.15 & 5.05 & 5.90 &	  &	      \\ \hline
Nonrelativistic constituent \cite{LlanesEstrada:2005jf}       
& 3.49 & 3.92 & 5.15 & 6.14 &	  &	      \\ \hline
Wilson loop \cite{Kaidalov:1999yd}                             
& 3.49 & 4.03 &      &        &        &       \\ \hline
Vacuum correlator \cite{Kaidalov:2005kz}                       
& 3.02 & 3.49 & 4.18 & 4.96   &        &       \\ \hline
Vacuum correlator \cite{Kaidalov:2005kz}                       
& 3.32 & 3.83 & 4.59 & 5.25   &        &       \\ \hline
Semirelativistic potencial \cite{Mathieu:2008pb}               
& 3.99 & 4.16 & 5.26 &        &        &       \\ \hline
Anisotropic lattice \cite{Chen:2005mg}                         
& 3.83 & 4.20 &      &        &        &       \\ \hline
Isotropic lattice \cite{Meyer:2004jc, Meyer:2004gx}            
& 3.24 & 4.33 &      &        &        &       \\ \hline
\end{tabular}
%\end{ruledtabular}
\end{table}

\section{Odderon Regge trajectories in the Hardwall model}

Taking the data for odd spin glueball masses obtained in the previous section we are going to build up the Regge trajectories for the odderon, using linear regression.

For the Dirichlet boundary condition and the set of states, $1^{--}$, $3^{--}$, $5^{--}$, $7^{--}$, $9^{--}$, $11^{--}$, we find the following Regge trajectory: 
\begin{equation}\label{reg_dir}
J^{\left\{ 1-11 \right\}}_{Dir.} (m^2) = -(0.83 \pm 0.40) + (0.22 \pm 0.01)m^2\,.
\end{equation}

\noindent The errors for the slope and linear coefficient come from the linear fit. The plot relative to this trajectory can be seen in Fig.  $1$. This result is in  agreement with that found in \cite{LlanesEstrada:2005jf}, with the relativistic many-body Hamiltonian formulation, described by Eq.  (\ref{l1}).

\begin{figure}[t] \label{dirg}
  \centering
  \includegraphics[bb=0 0 403 315,width=8cm,height=6.25cm,keepaspectratio]{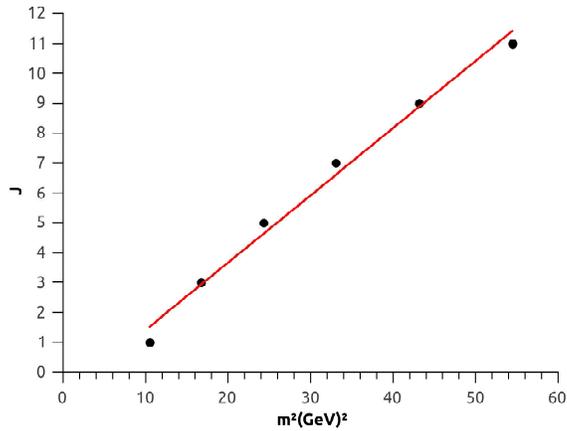} 
\caption{Glueball masses (dots) for the states $1^{--}$, $3^{--}$, $5^{--}$, $7^{--}$, $9^{--}$, $11^{--}$ from the holographic hardwall model using Dirichlet boundary condition, Eqs. (\ref{dir}) and (\ref{sdi}). We also plot an approximate linear Regge trajectory, corresponding to Eq. (\ref{reg_dir}), representing the odderon.}
\end{figure}

It has been argued in Ref. \cite{LlanesEstrada:2005jf} that the state $1^{--}$ might not be part of the spectrum of the odderon. To test this possibility we also consider another set of states, $3^{--}$, $5^{--}$, $7^{--}$, $9^{--}$, for which we obtain the following Regge trajectory:
\begin{equation}\label{reg_dir_1}
J^{\left\{ 3-9 \right\}}_{Dir.} (m^2) = -(0.63 \pm 0.31) + (0.23 \pm 0.01)m^2 .
\end{equation}

\noindent This result is also consistent with the Regge trajectory for odderon, Eq. (\ref{l1}).

Now using the Neumann boundary condition and the set of states, $1^{--}, 3^{--}, 5^{--}, 7^{--}$, $9^{--}, 11^{--}$, we find the following Regge trajectory:
\begin{equation}\label{reg_neu}
J^{\left\{ 1-11 \right\}}_{Neu.} (m^2) = -(0.29 \pm 0.42) + (0.18 \pm 0.01)m^2 .
\end{equation}

\noindent The plot relative to this trajectory can be seen in Fig. 2.

\begin{figure}[t]\label{neug} 
% float placement: (h)ere, page (t)op, page (b)ottom, other (p)age
  \centering
  % file name: D:/Documentos/Alunos_pos/Eduardo/neun.eps
  \includegraphics[bb=0 0 403 315,width=8cm,height=6.25cm,keepaspectratio]{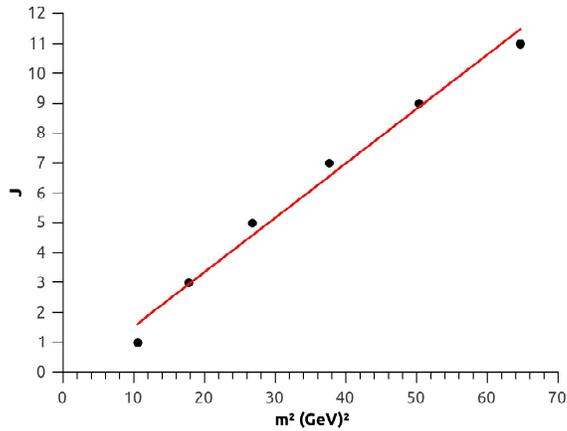} 
\caption{Glueball masses (dots) for the states $1^{--}$, $3^{--}$, $5^{--}$, $7^{--}$, $9^{--}$, $11^{--}$ from the holographic hardwall model using Neumann boundary condition, Eqs. (\ref{neu}) and (\ref{neu2}). We also plot an approximate linear Regge trajectory, corresponding to Eq. (\ref{reg_neu}), representing the odderon.}
\end{figure}

We also consider here the possibility of excluding the state $1^{--}$ from the spectrum of the odderon. For the set of states $3^{--}, 5^{--}, 7^{--}, 9^{--}, 11^{--}$, we find the following Regge trajectory: 
\begin{equation}\label{reg_neu_2}
J^{\left\{ 3-11 \right\}}_{Neu.} (m^2) = (0.34 \pm 0.37) + (0.17 \pm 0.01)m^2 .
\end{equation}

\noindent This result is in  agreement with that found in \cite{LlanesEstrada:2005jf}, with the nonrelativistic constituent model, Eq. (\ref{l2}).

\section{conclusions}

In this work we obtained odd spin glueball masses in the sector $P=C=-1$ using the holographic hardwall model with Dirichlet and Neumann boundary conditions. These glueball masses lie in approximate linear Regge trajectories compatible with results for the odderon, both in the relativistic many-body as well in the nonrelativistic constituent models presented in Ref. \cite{LlanesEstrada:2005jf}. The present analysis gives support to the conclusion of Ref. \cite{LlanesEstrada:2005jf} about the general properties of the odderon Regge trajectories, i.e., a low intercept and a slope similar to that of the Pomeron.

Some aspects of the holographic approach for the odderon Regge trajectories remain open. In our approach, we used Dirichlet and Neumann boundary conditions in the hardwall model obtaining results compatible with those of Ref. \cite{LlanesEstrada:2005jf}.  The hardwall model was used before to obtain the Regge trajectory for the Pomeron in Ref. \cite{BoschiFilho:2005yh}. In that work it was possible to conclude that the Neumann boundary condition is more appropriate than the Dirichlet boundary condition by comparison with experimental data. Here in this work, it is not possible to reach a similar conclusion about boundary conditions because there is no clear experimental data for the odderon Regge trajectories.

Another open question in the odderon Regge trajectories regards the state $1^{--}$. It was argued in Ref. \cite{LlanesEstrada:2005jf} that the glueball state $1^{--}$ does not belong to the odderon Regge trajectory. However, our analysis is not conclusive regarding this point since we have found trajectories compatible with odderon including the state $1^{--}$ [Eqs. (\ref{reg_dir}) and (\ref{reg_neu})] as well as excluding it [Eqs. (\ref{reg_dir_1}) and (\ref{reg_neu_2})]. 

As a final remark, let us comment on our choice for the holographic model to obtain glueball masses and the odderon Regge trajectories. This model is very interesting since masses can be obtained from the zeros of the corresponding Bessel functions. However, it is well known that a holographic hardwall model leads to asymptotic nonlinear Regge trajectories for very high states. Nevertheless, for light states, as discussed in this work, approximate linear Regge trajectories were found. In this regard, it will be interesting to investigate the glueball masses in the $P = C = -1 $ sector within other holographic approaches, such as the softwall model \cite{Karch:2006pv, Colangelo:2007pt}, which is known to provide linear Regge trajectories. We leave this study for future work.

\begin{acknowledgments}

We would like to thank Felipe J. Llanes-Estrada for interesting discussions and Nelson R. F. Braga for a careful reading of the manuscript. The authors are partially supported by CAPES, CNPq and FAPERJ, Brazilian agencies.
 
\end{acknowledgments}

\end{document}